# Detection of infalling hydrogen in transfer between the interacting galaxies NGC 5426 and NGC 5427.

Joan Font (1,2), John E. Beckman (1,2,3), Margarita Rosado (4), Benoît Epinat (5,6) Kambiz Fathi (1,7,8), Olivier Hernandez(9), Claude Carignan (9), Leonel Gutiérrez (1,10), Monica Relaño (11), Javier Blasco-Herrera (7), Isaura Fuentes-Carrera (12).

1. Instituto de Astrofísica de Canarias,c/ Vía Láctea, s/n, E38205, La Laguna, Tenerife, Spain: jfont@iac.es, jeb@iac.es.
2. Departamento de Astrofísica. Universidad de La Laguna, Tenerife, Spain.
3. Consejo Superior de Investigaciones Científicas, Spain.
4. Instituto de Astronomía, Universidad Nacional Autónoma de México, Apdo. Postal 70-264, 04510, México D.F., México: margarit@astroscu.unam.mx.
5. Laboratoire d'Astrophysique de Toulouse-Tarbes. Université de Toulouse, CNRS, 14 Avenue Édouard Berlin, F-31400, Toulouse, France: benoit.epinat@ast.obs-mip.fr.
6. CNRS, IRAP, 9 Av. Colonel Roche, BP 44346, 31028, Toulouse, Cedex 4, France.
7. Department of Astronomy, University of Stockholm, AlbaNova, 10691, Stockholm, Sweden: kambiz@astro.su.se, javier@astro.su.se.
8. Oscar Klein Centre for Cosmoparticle Physics, Stockholm University, 10691 Stockholm, Sweden.
9. Département de Physique et Astronomie, Université de Montreal, CP. 6128, Succ. Centre ville, Montréal, QC H3C 3J7, Canada: hernandez@astro.umontreal.ca, carignan@astro.umontreal.ca.
10. Instituto de Astronomía, Universidad Nacional Autónoma de México, Apdo. Postal 877, 22800, Ensenada B.C., México: leonel@astrosen.unam.mx.
11. Universidad de Granada, Av. Severo Ochoa, s/n, 18071, Granada, Spain: mrelano@ugr.es.
12. Departamento de Física. Escuela Superior de Física y Matemáticas. IPN, U.P. Adolfo Lopez Mateos, C.P. 07738, México City, México: isaura@esfm.ipn.mx.

ABSTRACT

Using velocity tagging we have detected hydrogen from NGC 5426 falling onto its interacting partner NGC 5427. Our observations, with the GHaFaS Fabry-Perot spectrometer, produced maps of the two galaxies in Hα surface brightness and radial velocity. We found emission with the range of velocities associated with NGC 5426 along lines of sight apparently emanating from NGC 5427, superposed on the velocity map of the latter. After excluding instrumental effects we assign the anomalous emission to gas pulled from NGC 5426 during its passage close to NGC 5427. Its distribution, more intense between the arms and just outside the disk of NGC 5427, and weak, or absent, in the arms, suggests that the infalling gas is behind the disk., ionized by Lyman continuum photons escaping from NGC 5427. Modeling this, we estimate the distances of these gas clouds- behind the plane: a few hundred pc to a few kpc. We also estimate the mass of the infalling (ionized plus neutral) gas, finding an infall rate of 10 $M_\odot$ per year, consistent with the high measured SFR across the disk of NGC 5427 and with the detected circumnuclear galactic wind.

*Subject headings*: galaxies: individual (NGC5426, NGC 5427, Arp271) – galaxies: kinematics and dynamics – galaxies: interactions – galaxies: spiral – galaxies: ISM.

1. INTRODUCTION.

As the mean density of the universe in earlier epochs was higher than that in the local universe, merging is known have played a key role in the evolution of the galaxies we see today, so to study the process of merging holds great interest, but this process is best studied locally, using high spatial resolution and S:N observations. The Atlas of Peculiar Galaxies by Arp (1966) in which many are interacting is a major starting point for such studies. Toomre and Toomre´s (1972) early simulation models accounted for many features in this atlas, and indicated two key features of mergers: material driven towards galaxy centers, and driven into crowded orbits across the disks. Using stars plus gas, later simulations indicated enhanced star formation across the disks, and notably in circumnuclear zones. In a Letter we cannot review this and refer the reader to the review by Struck (2006), on the observational side, and to the conference proceedings edited by Smith et al. (2010) for an overview.

Simulations of interactions incorporate kinematic as well as structural information, but there have not been many detailed kinematic optical observations of interacting or merging galaxies, and most have been confined to limited zones using fiber-fed spectrographs (many kinematic studies based on radio observations of gas in interacting galaxies have been reported, but at intermediate angular resolution, e.g. Van der Hulst 1979, Vollmer et al. 2005). Here we show the potential of this type of observations for revealing details about the interaction process, over complete galaxy discs, using Fabry-Perot spectroscopy giving two dimensional kinematics across a relatively large field. Fabry-Perot observations of interacting galaxies have been reported previously; Rampazzo et al. (2005) studied five such systems, while Fuentes-Carrera et al. (2004) presented observations of Arp 271, the subject of this letter, and the same group have carried out more recent work in this field (see e.g. Repetto et al., 2010). In this letter we report the results of observations of Arp 271 (NGC 5426+NGC 5427) using a Fabry-Perot spectrometer.

2. THE OBSERVATIONS.

The observations were made on the night of July $5^{th}$-$6^{th}$ 2007, with GH$\alpha$FaS, a Fabry-Perot interferometer-spectrometer (Hernandez et al. 2008) at the Nasmyth focus of the 4.2m William Herschel Telescope, Observatorio del Roque de los Muchachos, La Palma. The Nasmyth field size is restricted to 3.5 arcmin x 3.5 arcmin, so that although we made a complete map of NGC 5427, only some 60% of the bright galaxy disk of NGC 5426 was included. After correction, velocity calibration and phase adjustment we obtained a "data cube" from which we could extract maps of the interacting galaxies in surface brightness, in velocity, and in velocity dispersion, from the H$\alpha$ emitted by the HII regions and the diffuse gas. In Figure 1(a) we show the surface brightness map, compared with a broad-band image (GSA). The HII regions, principally in the arms and in the centre region of NGC 5427, are prominent, and similarly for NGC 5426. In Figure 1(b) we show the velocity map of NGC 5426 derived by plotting the peaks of the emission lines. In this map towards the top of the field, there is a zone with a range of velocities coincident with that of NGC 5426, but with morphology comparable to that of NGC 5427. This section of the map was made using a set of "anomalous" emission peaks, found when examining the spectra of NGC 5427, and it forms the main subject of the present article. In Figure 2 we show the velocity map of NGC 5427 together with the map of the anomalous peaks (see the figure caption for details).

We then derived the rotation curves of both galaxies. There are non-rotational velocity components, but their projections close to the major axes, where we measured these curves, are minimal. In Figure 3 we plot both curves, separated along the distance axis by the projected distance between the galaxy centers. The curve for NGC 5427 stops at short distances from the nucleus on either side: there is no strong emission beyond this radius, but the curve for NGC 5426 extends much further on the north side (the south is cut off by the edge of the field). The points further from its nucleus than 18.3 kpc are obtained from the spots overlapping with NGC 5427, corresponding to the prolongation of the optical bridge, and including the anomalous map. The curve gives a global picture of the velocity distributions but does not take into account small scale departures from rotational symmetry. Only major systematic departures can be detected. The curve for NGC 5426 shows strong apparent oscillations where it passes the kinematic centre of NGC 5427, but then settles as a declining projection of the main curve for NGC 5426. We will interpret this curve with the help of the map of the anomalous velocity component, in section 3.

3. THE "ANOMALOUS" VELOCITY MAP, AND ITS INTERPRETATION.

Before interpreting the anomalous velocity map we excluded the possibility of an instrumental artifact. Fabry-Perot systems produce ghost images of low intensity, displaced from the main image. The optics are designed to form the ghost diametrically opposite the true image, so that the anomalous emission might be a ghost of NGC 5426. However this is not so, for three reasons, relying on measurements on ghost images of other objects from GHaFaS. Firstly the morphology of the anomalous emission is not similar to that of NGC 5426, as expected for a ghost. Secondly its luminosity is an order of magnitude greater than we find in our ghosts (more than 20% of the luminosity of NGC 5426, rather than the 2% upper limit for ghosts), and thirdly the wavelength range emitted by the anomalous emission is different from that of the strongest emitting parts of NGC 5426, rather than being the same, as for a ghost. As NGC 5427 is much more luminous than NGC 5426 (see section 4 below) if the anomalous emission were from a ghost, we would expect much stronger anomalous emission due to a ghost of NGC 5427 in the general zone of NGC 5426, and this is absent.

The clue to the nature of the anomalous emission comes from a careful look at the morphology. This emission comes mainly from the zones between the spiral arms, or close to them but just off the edge of the disk of NGC 5427. There is a negative correlation between the line strengths of the anomalous component and the normal component. We illustrate this in Figure 2, with six representative spectra, which come from the points on the image shown. The spectrum from the arm center shows a strong "normal" line (i.e. at the velocity of NGC 5427) and no anomalous line, while the spectrum from just off the edge of the disk shows a strong anomalous line and no normal line. The other spectra show both lines, with the stronger normal lines accompanied by weaker anomalous lines, and vice versa.

These relationships suggest that the anomalous emission comes from behind the disk of NGC 5427, so we detect it only where the dust extinction in the disk is low, in the interarm zones and off the disk edge. This interpretation receives support from previous observations. Arp 271 has been observed at many wavelengths with published images in the visible (Arp 1966, also see APOD 2008), radio (Meyer et al. 2004) near and mid infrared with Spitzer (Smith et al. 2007), and in the UV with GALEX (Smith et al. 2010). Its kinematics were analyzed from previous Fabry-Perot observations by Fuentes-Carrera et al. (2004) showing that the systemic velocities of the two galaxies differ by less than 200 km/sec, and suggesting that the mutual orbits of the two galaxies have taken NGC 5426 behind NGC 5427, from which is has emerged. A "bridge" of material drawn gravitationally from NGC 5426 by NGC 5427 is visible in optical and infrared images, including those from Spitzer in the 3.5 μm and 8 μm bands, (Smith et al. 2007).

As the anomalous component has a velocity range comparable to NGC 5426, it seems reasonable to infer that it comes from material related kinematically to NGC 5426, but if it is hydrogen detached from the disk, in the process of capture by NGC 5427 it is unlikely to have sufficient density to produce massive stars which could ionize it internally. To explain the anomalous emission we found a clue from the high velocity clouds (HVC´s) around the Milky Way, which have been detected in Hα at the same velocity as the previously detected 21cm HI emission (e.g. Tufte et al. 1993 and Putman et al. 2003). Their Hα surface brightness has been used to estimate their distances above the Galactic plane (Putman et al. 2003), using models by Bland-Hawthorn & Maloney (1997) which include escaping Lyman continuum (Lyc) photons from the disk. The emission from an HVC depends on its distance from the sources of Lyc photons, i.e. from the HII regions. Using this scenario we can understand the pattern of the anomalous emitters towards NGC 5427. The brightness of an emitter depends on its distance from the sources of Lyc photons. The pattern reflects the distribution of the Lyc sources, i.e. it is a pseudo-image of the HII regions of NGC 5427 in the ionized infalling gas from NGC 5426. This pseudo-image is modulated by the dust distribution as it passes through the disk from behind, leaving the pattern we observe via our technique of velocity tagging.

We note a further point in our interpretation of the observations. In Figure 3 the "rotation curve" of NGC 5426 shows a considerable disturbance in the range of galactocentric distance between some 19 and 26 kpc. At smaller radii it behaves in the conventional almost invariant manner, and at larger radii shows a steady decline, but in this radial range it shows strong variations. We interpret this as due to the presence of outflow from the centre of NGC 5427, such that our measured lines are blends with the outflow emission. The peak velocity of this outflow is at 22.5 kpc from the projected center of NGC 5426, but more significantly coincides with the centre of NGC 5427. The outflow velocity, measured with respect to the systemic velocity of NGC 5427, is just under 400 km/sec, consistent with velocities measured in galactic winds.

4. DISTANCES OF THE INFALLING CLOUDS FROM THE PLANE OF NGC 5427, AND THE INFLOW RATE.

Our scenario lets us estimate the distance of any approaching cloud behind the plane of NGC 5427, assuming case B emission (Osterbrock 1989) from the HII regions of the galaxy and from the approaching clouds, and a constant escape fraction of Lyc photons from the regions. Using a flux-calibrated map of the Hα luminosities of the HII regions obtained at the 2.2m telescope at San Pedro Mártir Observatory (UNAM, Ensenada, México) we calculated the Lyc field in the 3d space behind NGC 5427, using a distance to the galaxy of 34.8 Mpc, from our recession velocity measurement, with a Hubble constant of 71 km·sec$^{-1}$·Mpc$^{-1}$. From the distance of an emission spot of the ionized cloud from the galaxy centre projected onto the plane we could compute its perpendicular distance behind the plane from its measured surface brightness. These distances are upper limits if we do not allow for dust extinction in the semi-transparent parts of the galaxy plane, but beyond the disc edge this effect is very small. From Galactic data (Bland-Hawthorn & Maloney 1999) we might have used a 5% escape fraction for the Lyc photons, but we consider this too low, as there is indirect evidence that the escape fraction rises with HII region luminosity (Zurita et al. 2002, Gutiérrez & Beckman 2010). From our calibrated image we determined that the Hα luminosities of the HII regions in NGC 5427 are half an order of magnitude greater than those of the most luminous Galactic regions, a symptom of the enhanced SFR due to the interaction and gas infall. From the distance estimates of the clouds behind the plane of NGC 5427, obtained using an optical depth 0.5 for clouds within the disk radius, and a Lyc escape factor of 20% we produced a 3d model of their distribution, which is schematically illustrated in Fig. 4. This distribution is quantitatively reasonable, but cannot be treated as precise.

We estimated the masses of ionized gas in the clouds, first deriving their emission measures (see Osterbrock, 1989) and then electron densities and ionized gas column densities, following Bland-Hawthorn and Maloney (1997, 1999) as applied by Putman et al (2003) for Galactic HVC´s. We assumed sphericity to first order, and estimated the cloud radii given the distance to NGC 5427. Masses of ionized gas in individual clouds were of order $5\times10^5$ M$_\odot$, and an integrated value for all the detected clouds is $5\times10^7$ M$_\odot$.

Using the distances of the clouds from the plane of NGC 5427, and their approach velocities we estimated an accretion rate of order 1 M$_\odot$ of ionized hydrogen per year. The Galactic observers (Tufte et al. 1998, Putman et al. 2003) could measure both the ionized and the neutral masses of the same HVC´s, giving a factor of 10 between them. Using this value we derive a mass inflow to NGC 5427 of order 10 M$_\odot$ per year, distributed widely across the disk. This is consistent with our measurement of the Hα luminosity of NGC 5427: 41.9 dex (log units of erg s$^{-1}$), well above the values normally found for Milky Way mass galaxies. Zurita et al. (2000) found a mean for 5 "normal" galaxies of 41.4 dex, and 42.1 dex for the starburst galaxy NGC 7479; we can also compare this with our measurement of 41.4 dex for NGC 5426.

The enhanced SFR is distributed across the disc of NGC 5427, stimulated by the infall process (Casuso et al. 2006), and fuelled by the infalling gas. Inflow to the centre within the plane of NGC 5427 enhances the local SFR near the nucleus, producing the observed galactic wind. The system is initiating its merger, and the relative weakness of the initial phases of the interaction has allowed us to detect the details of the gas response rather clearly using velocity at the tagging agent. The observational picture presented here offers a newly examined scenario for modelers of galaxy-galaxy interactions.

ACKNOWLEDGMENTS

We thank Philippe Amram, Jean-Luc Gach, Philippe Balard, Jean Boulesteix, and Marie-Maude de Denus-Baillargeon, for invaluable technical and scientific inputs. Support was from project AYA2007-67625-CO2-01 of the Spanish Ministry of Science and Innovation, and project 3E 310386 of the Instituto de Astrofísica de Canarias. Based on observations from the William Herschel Telescope, Isaac Newton

Group of Telescopes, Observatorio del Roque de los Muchachos, Insituto de Astrofísica de Canarias, La Palma, Spain, and the 2.2 m telescope at the Observatorio de San Pedro Martir, UNAM, Ensenada, México, and the Gemini Observatory, operated by AURA Inc. under a cooperative agreement with the NSF on behalf of the Gemini partnership. We warmly thank the anonymous referee for comments which led to improvements in the article.

FIGURES

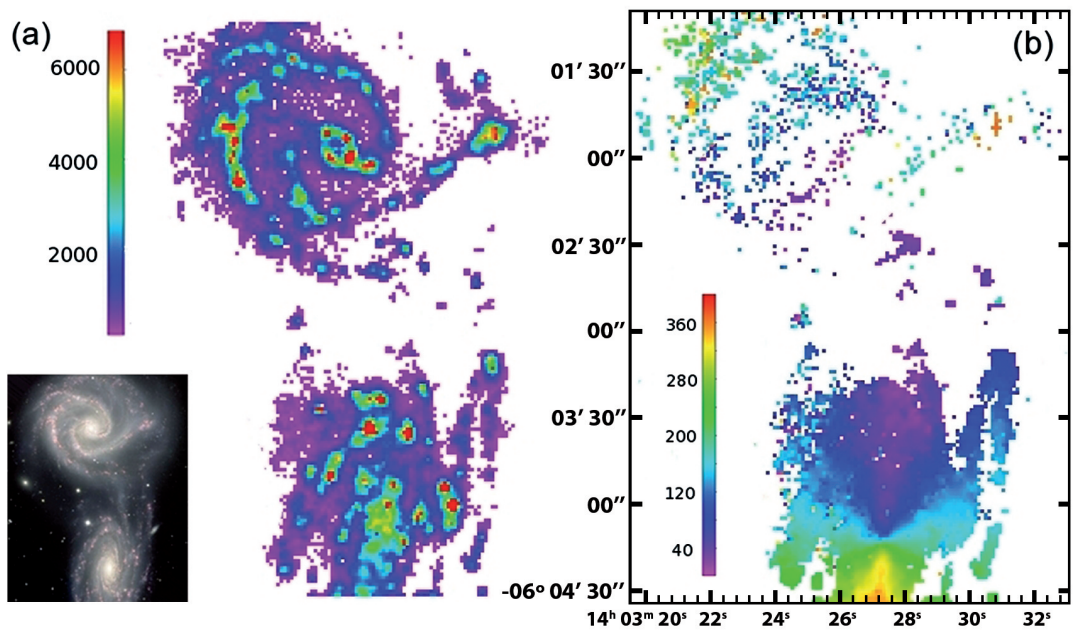

**Figure 1.** (a) Comparison of a composite visible image of Arp 271 (bottom left) taken from the Gemini Science Archive with an image in Hα surface brightness taken with GHαFaS; NGC 5426 is at the bottom of the image and NGC 5427 at the top. The color scale is in relative units of surface brightness. The ionized gas in the main star forming regions (HII regions) stands out clearly, concentrated mainly in the arms and round the nucleus.
(b) Velocity map of NGC 5426 alone (velocity key to the left in km s$^{-1}$), made by transforming the observed Hα wavelengths across the galaxy to velocities, via their Doppler shifts. The zero point was chosen so that the velocity is continuously positive across the figure. The upper section of the figure is gas from NGC 5426 pulled towards NGC 5427, and emitting Hα stimulated by Lyman continuum photons escaping from NGC 5427, it is an "ionization image" of NGC 5427 in the infalling gas (see text for details).

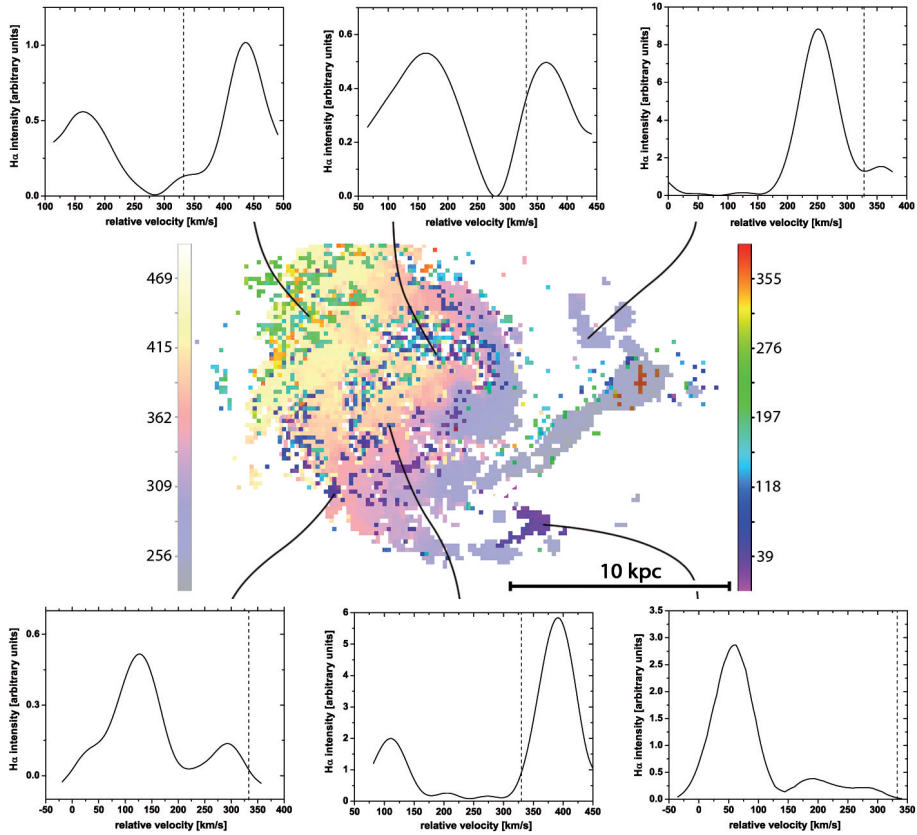

**Figure 2**. Composition of the velocity map of NGC 5427 -lighter colors- and that of the infalling gas -darker colors-; (the latter is displayed more clearly in the upper part of panel (b)). Color coded velocity tables for each component are shown at either side of the panel. Above and below the map we show Hα line profiles at the places indicated. Vertical line: systemic velocity of NGC 5427 Vertical In each spectrum the lower velocity peak is from the infalling gas component of NGC 5426, and the other peak from the disk of NGC 5427. The chosen cases show how the infalling gas is detected between the arms and outside the disk.

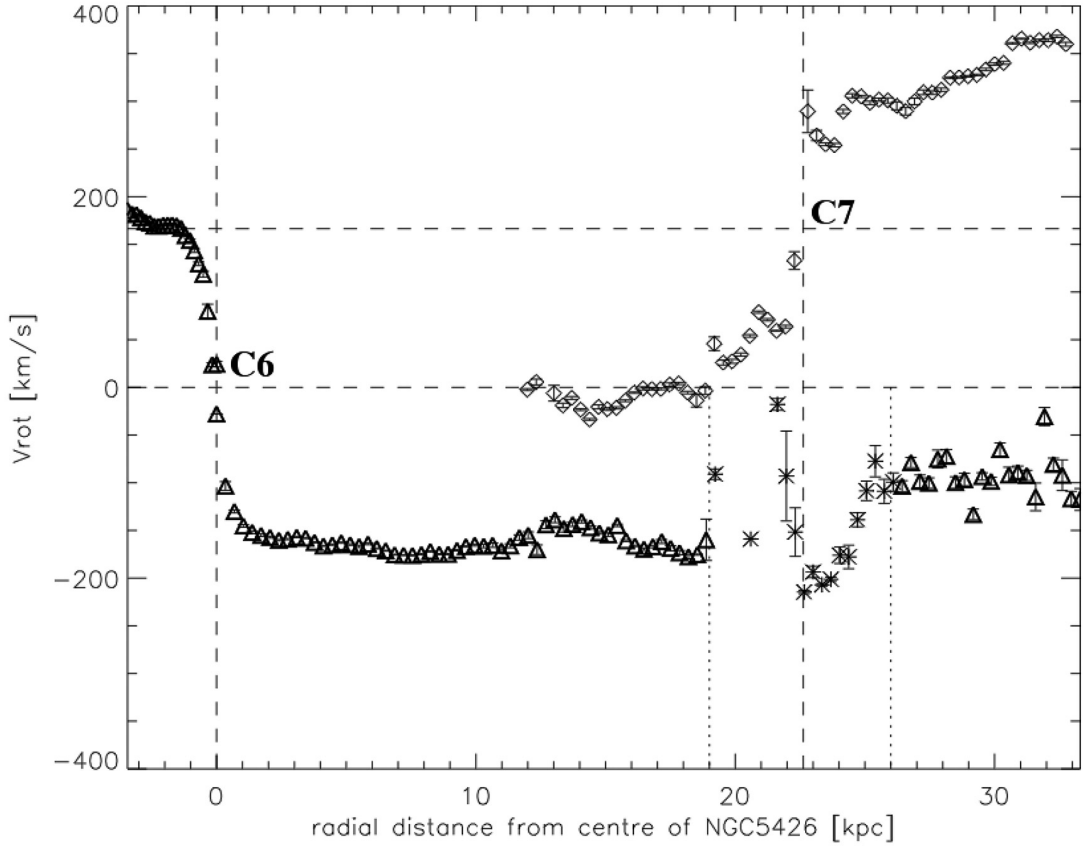

**Figure 3.** Rotation curves of NGC 5426 (triangles) and NGC 5427 (diamonds). The center position of NGC 5426 and its systemic velocity are set set to zero for reference. The curve for NGC 5427 is offset by 22.6 kpc, the separation of the galaxy centers projected onto the major axis of NGC 5426 and by 165.5 km/sec, the difference between their systemic velocities. Their zero points are labeled C6 and C7 respectively. Vertical dotted lines limit the zone where the rotation curve of NGC 5426 is affected, and partly replaced, by emission from the galactic wind from the center of NGC 5427. These velocity points are marked as asterisks.

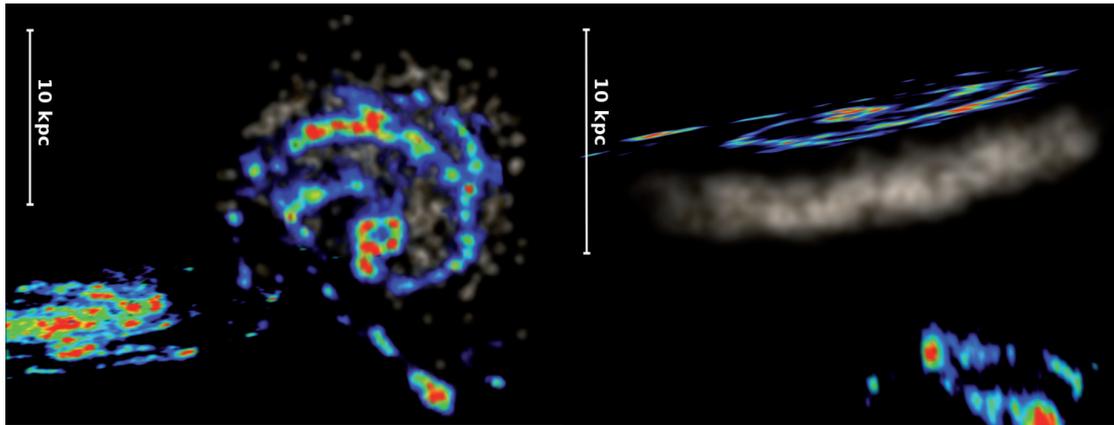

**Figure 4.** Pictorial representations of Arp 271 including the clouds falling towards NGC 5427. For optimum visualization we provide two views a "sky view" (left hand panel) showing the clouds as they are detected behind the plane, (North is to the right, East is up), and an edge-on view (right panel) indicating the estimated range of distances of the clouds from the plane.